\documentclass[onecolumn,prd,aps,12pt]{revtex4}%
\usepackage{amsmath}
\usepackage{amsfonts}
\usepackage{amssymb}
\usepackage{graphicx}%
\providecommand{\U}[1]{\protect\rule{.1in}{.1in}}
\begin{document}
\title{Metamaterials from modified CPT-odd standard model extension and minimum length}
\author{T. Prudencio$^{1}$\thanks{email: prudencio.thiago@ufma.br}, H. Belich$^{2}%
$\thanks{email: belichjr@gmail.com}, H. L. C. Louzada$^{2}$\thanks{email:
haofisica@bol.com.br.}}
\affiliation{$^{1}${\small Coordination of Science \& Technology (CCCT-BICT), Federal
University of Maranh\~ao (UFMA), 65080-805, S\~ao Lu\'is-MA, Brazil.}}
\affiliation{$^{2}${\small Departamento de F\'{\i}sica e Qu\'{\i}mica, Federal University
of Esp\'{\i}rito Santo (UFES), 29060-900, Vit\'{o}ria, ES, Brazil.}}
\affiliation{}
\date{\today}

\begin{abstract}
Here we discuss the standard model extension in the presence of CPT-odd
Lorentz violation (LV) sector and of a deformed Heisenberg algebra that leads
to a non-commutative theory with minimum length (ML). We derive the set of
Maxwell equations emerging from this theory and considered the consequences
with respect to the usual effects of electromagnetic waves and material media.
We then considered the set of modified equations in material media and
investigate the metamaterial behaviour as a consequence of LV and ML. We show
that a negative index refraction can be derived from the presence of a
non-commutativity suitably tuned by the $\beta$ parameter, while in the
presence of LV, we obtained the set of modified Maxwell equation in terms of
the corresponding material fields with terms depending explicitly from the
terms of interaction between the material fields depending on
non-commutativity with the background field due to CPT-odd LV. We conclude
that a new set of metamaterials can be derived as a consequence of CPT-odd LV
and non-commutativity with minimum length.

\end{abstract}
\maketitle

\section{Introduction}

The induction of Spontaneous Lorentz Violation (SLV) symmetry by the presence
of a tensorial background is one of the possible scenarios in the
non-homogeneous spacetime that could to explain Planck scale physics and dark
physics scenarios. The residues of quantum gravity and metric fluctuations where
the standard model (SM) has achieved its validity limit, brought a more
general picture represented by Standard Model Extension. The Standard Model
Extended (SME) is an effective field theory obtained from SM (the model that
uniquely describes the interactions that govern elementary particles, namely
electromagnetic, weak nuclear and strong nuclear interactions, but does not
incorporate the Gravitational) by the addition of terms that incorporate the
violations of the Lorentz and CPT symmetry. In this scenario, Lorentz symmetry
violation (LV) is a natural phenomena in high energy scales, inducing both
spontaneous Lorentz symmetry violation (SLV) caused by a tensorial background
in one side, and the breaking made by generalization of uncertainty principle,
that is associated to a non-commutative geometry. Kosteleck\'{y} and Samuel
\cite{sam} proposed the idea of SLV when they initiated investigations of a
possible physics beyond the Standard Model (SM). They have suggested that SLV
could occur in a scenario of string field theory by means of non-scalar fields
(vacuum of fields that have a tensor nature) by the presence of a non-scalar
background \cite{col,coll2}. As a consequence, the prediction involves the
assumption that in a more fundamental theory signals could be emitted from
more fundamental fields by SLV symmetry. SME keeps the gauge invariance,
conservation of energy and momentum and the covariance under observer
rotations and boosts \cite{ens}. In this context, it is well-known that the
presence of terms of LV symmetry imposes at least one privileged direction in
the spacetime. In recent years, studies of the LV symmetry have been made in
several different contexts
\cite{h1,h2,e1,e2,e3,e4,e5,e6,e6a,e7,rasb1,w,tensor1,tensor2,geom1,geom2,geom3,bb2,bb4,book,l1,g1,g2,g3}%
.

On the other hand, non-commutative geometry (NCG) was developed some decades
ago by A. Connes \cite{33} and it was realized that the NCG would be a scheme
to extend the SM in several forms \cite{necra}. In particular, it appears
naturally in string theory scenarios \cite{34, 44}, that should lead to
effective theories describing scenarios beyond SM and that recover at low
energy limits known physical results from SM.

One possible way to explore the implementation of NCG theories is by the
deformation of the Heisenberg algebra. In a modified Heisenberg algebra, by
adding certain small corrections to the canonical commutation relations, it
leads, as shown by A. Kempf and contributors \cite{K1,K2,K3,K4,K5}, to the
minimum uncertainty in the position measurement, $\Delta x_{0},$ called
minimum length. The existence of this minimum length was also suggested by
quantum gravity and string theory \cite{DJ,MM,EW}.

Quesne and Tkachuk have introduced recently a Lorentz covariant deformed
algebra in a quantized $D+1$-dimension \cite{QT,QT2}, whose generalized
commutation relations can be written:
\begin{align}
\frac{1}{i\hbar}\left[  X^{\mu},P^{\nu}\right]   &  =(-1+\beta P^{2})\eta
^{\mu\nu}+\beta^{^{\prime}}P^{\mu}P^{\nu}\\
\frac{1}{i\hbar}[P^{\mu},P^{\nu}]  &  =0,\label{es1}\\
\frac{1}{i\hbar}\left[  X^{\mu},X^{\nu}\right]   &  =\frac{[(2\beta
-\beta^{^{\prime}})-(2\beta^{2}+\beta\beta^{^{\prime}})P^{2}](P^{\mu}X^{\nu
}-P^{\nu}X^{\mu})}{(1-\beta P^{2})},
\end{align}
where $P^{2}=P_{\rho}P^{\rho}$, $\mu,\nu,\rho=0,1,\cdots,D$, $\quad{\eta
_{\mu\nu}}=\eta^{\mu\nu}=diag(1,-1,-1,\cdots,-1)$, and $\beta,\beta^{^{\prime
}}>0$ are deformation parameters. From uncertainty relation, a minimum length
can be achieved for these commutation relations, that is given by
\[
\left(  \delta X^{i}\right)  _{0}=\hbar\sqrt{(D\beta+\beta^{^{\prime}})\left[
1-\beta\left\langle (P^{0})^{2}\right\rangle \right]  }\quad,\quad\forall
i\in\left\{  1,\cdots,D\right\}  .
\]
The representation algebra \cite{VM}, satisfying the above commutation
relations at first order in $\beta,\beta^{^{\prime}}$\ are given by:%
\begin{align}
X^{\mu}  &  =x^{\mu}-\frac{(2\beta-\beta^{^{\prime}})}{4}(x^{\mu}p^{2}%
+p^{2}x^{\mu}),\label{es2}\\
P^{\mu}  &  =\left(  1-\frac{\beta^{^{\prime}}}{2}p^{2}\right)  p^{\mu},
\end{align}
where $p^{2}=p_{\rho}p^{\rho}$.
The corresponding the position and momentum operators $x^{\mu}$ and $p^{\mu
}=i\hbar{\partial}^{\mu}$ will lead to the equivalent relations
\begin{align}
X^{\mu}  &  =x^{\mu}+\frac{(2\beta-\beta^{^{\prime}})\hbar^{2}}{4}(x^{\mu}%
\Box+\Box x^{\mu}),\label{es3}\\
P^{\mu}  &  =\left(  1+\frac{\beta^{^{\prime}}}{2}\hbar^{2}\Box\right)
i\hbar{\partial}^{\mu},
\end{align}
where $\Box=\partial_{\mu}\partial^{\mu}$.

At second order in $\hbar$, this reduces to
\begin{eqnarray}
X^{\mu} &=& x^{\mu}\left(  1+\frac{(2\beta-\beta^{^{\prime}}%
)\hbar^{2}}{4}\Box\right)  ,\label{uk}\\
P^{\mu} &=& i\hbar{\partial}^{\mu} \label{uk2}
\end{eqnarray}
and up to third order in $\hbar$,
\begin{align}
X^{\mu} &  =x^{\mu}\left(  1+\frac{(2\beta-\beta^{^{\prime}}%
)\hbar^{2}}{4}\Box\right)  ,\label{ukq}\\
P^{\mu} &  =\left(  1+\beta\hbar^{2}\Box\right)  i\hbar{\partial}^{\mu
},\label{uk2q}%
\end{align}
We also can consider the case $\beta^{^{\prime}}=2\beta$, where
\begin{align}
X^{\mu} &  =x^{\mu},\label{es3q}\\
P^{\mu} &  =\left(  1+\beta\hbar^{2}\Box\right)  i\hbar{\partial}^{\mu
},\label{es33}%
\end{align}

As a consequence, taking $\beta^{^{\prime}}=2\beta$ and second order in
$\hbar$ we recover an usual canonical transformation. The hydrogen atom is one
of the simplest quantum systems that allows theoretical predictions of high
accuracy, and is well-studied experimentally offering the most precise amount
of measures \cite{SG}. There are many papers where the energy spectrum of the
hydrogen atom in the presence minimum length is calculated \cite{FB,SB,RA},
some of which have divergences in levels $s$ ($n=1$) \cite{SB}, and recently
it was studied the Stark effect with minimum length \cite{hao}.

We investigate the scenario of anisotropy in polarized electromagnetic waves
\cite{casana} generated by the presence of a Lorentz symmetry breaking term
defined by a term $\epsilon_{\mu\nu\alpha\beta}V^{\mu}A^{\nu}F^{\alpha\beta}$
\cite{jackiw}. The Lorentz symmetry violation is located in the fourvector
$V^{\mu}$ behavior. The effects of these anisotropies in the nature of vacuum
polarized electromagnetic waves is then discussed.

The structure of this paper is the following: Section II we consider the
CPT-odd sector of SME in the presence of a minimum length; section III we
consider the transformation of polarized electromagnetic wave under the effect
of a second order ML. Secion IV, we consider the non-commutativity in the
approximation to reduce the effect of transformation restricted to the
momentum sector, what leads to consequences in both electromagnetic waves and
in the presence of source terms. In section V, we discuss the metamaterial
behaviour with negative refractive index and in section VI we derive the set
of modified Maxwell equations in material media, depending on the CPT-odd
terms and ML. Finally, in section VII we leave our concluding remarks.

\section{The odd Gauge sector of the Standard Model Extended with minimum
length}

We start with the following CPT-odd gauge sector from SME \cite{col, coll2}
\begin{equation}
{\mathcal{L}}_{2N+1}=-\frac{1}{4\mu_{0}}F_{\mu\nu}F^{\mu\nu}-\frac{\chi}%
{4\mu_{0}}\epsilon_{\mu\nu\alpha\beta}V^{\mu}A^{\nu}F^{\alpha\beta}-A_{\mu
}J^{\mu},
\end{equation}
where $F^{\mu\nu}=\partial^{\mu}A^{\nu}-\partial^{\nu}A^{\mu}$ is the tensor
of the electromagnetic field constructed from the gauge field $A^{\mu}%
=(A^{0},\mathbf{A})$. The term $\epsilon_{\mu\nu\alpha\beta}V^{\mu}A^{\nu
}F^{\alpha\beta}$ leads to a possible scenario of anisotropy generated by LV
\cite{jackiw}, where Lorentz symmetry violation is located in the vector
$V^{\mu}$ behavior.

We will write this Lagrangian in the presence of a NCG given by (\ref{es3})
and (\ref{es33}), corresponding to the transformations
\begin{align}
x^{\mu}\rightarrow X^{\mu}  &  = x^{\mu},\\
\partial^{\mu}\rightarrow\nabla^{\mu}  &  =(1+\beta\hbar^{2}\Box)\partial
^{\mu}.\nonumber
\end{align}
Neglecting terms of non-linear order in $\beta$, we arrive at
\begin{align}
{\mathcal{L}}_{2N+1,NCG}  &  = {\mathcal{L}}_{2N+1} -\frac{1}{2\mu_{0}}%
\beta\hbar^{2}F_{\mu\nu}\Box F^{\mu\nu}-\frac{\chi}{4\mu_{0}}\beta\hbar
^{2}\epsilon_{\mu\nu\alpha\beta}V^{\mu}A^{\nu}\Box F^{\alpha\beta}.
\end{align}
The corresponding modified Maxwell equations of motion are given by
\begin{equation}
\label{em3}(1+2\beta\hbar^{2}\Box)\partial_{\nu}F^{\nu\mu}+\frac{\chi}%
{2}\epsilon^{\mu\lambda\alpha\beta}V_{\lambda}(1+\beta\hbar^{2}\Box
)F_{\alpha\beta}=\mu_{0}J^{\mu}.
\end{equation}
The usual Bianchi identity is left invariant
\begin{equation}
\label{em4}\partial_{\nu}F_{\alpha\beta}+\partial_{\alpha}F_{\beta\nu
}+\partial_{\beta}F_{\nu\alpha}=0.
\end{equation}

\section{Transformation up to second order in $\hbar$ and polarized plane
waves}

Let us consider a set of polarized plane wave solutions $\mathbf{E}(x^{\mu
})=\mathbf{E}_{0}\exp\left(  {k}_{\mu}x^{\mu}\right)  $ and $\mathbf{B}%
(x^{\mu})=\mathbf{B}_{0}\exp\left(  {k}_{\mu}x^{\mu}\right)  $. Under the
transformations (\ref{uk}) and (\ref{uk2}), these fields are changed to
following
\begin{align}
\mathbf{E}(x^{\mu}+\frac{(2\beta-\beta^{^{\prime}})\hbar^{2}}{4}(x^{\mu}%
\Box+\Box x^{\mu}))  &  =\mathbf{E}_{0}\exp\left(  {k}_{\mu}[x^{\mu}%
+\frac{(2\beta-\beta^{^{\prime}})\hbar^{2}}{4}(x^{\mu}\Box+\Box x^{\mu
})]\right)  ,\\
\mathbf{B}(x^{\mu}+\frac{(2\beta-\beta^{^{\prime}})\hbar^{2}}{4}(x^{\mu}%
\Box+\Box x^{\mu}))  &  =\mathbf{B}_{0}\exp\left(  {k}_{\mu}[x^{\mu}%
+\frac{(2\beta-\beta^{^{\prime}})\hbar^{2}}{4}(x^{\mu}\Box+\Box x^{\mu
})\right)  ,
\end{align}
and consequently the fields turn to behave as operators. We rewrite in the
form
\begin{align}
\mathbf{E}(x^{\mu}+\frac{(2\beta-\beta^{^{\prime}})\hbar^{2}}{4}(x^{\mu}%
\Box+\Box x^{\mu}))  &  =\mathbf{E}(x^{\mu})\exp\left(  {k}_{\mu}%
[\frac{(2\beta-\beta^{^{\prime}})\hbar^{2}}{4}(x^{\mu}\Box+\Box x^{\mu
})]\right)  ,\\
\mathbf{B}(x^{\mu}+\frac{(2\beta-\beta^{^{\prime}})\hbar^{2}}{4}(x^{\mu}%
\Box+\Box x^{\mu}))  &  =\mathbf{B}(x^{\mu})\exp\left(  {k}_{\mu}%
[\frac{(2\beta-\beta^{^{\prime}})\hbar^{2}}{4}(x^{\mu}\Box+\Box x^{\mu
})\right)  ,
\end{align}
Taking this up to the first order, we can rewrite
\begin{align}
\mathbf{E}(x^{\mu}+\frac{(2\beta-\beta^{^{\prime}})\hbar^{2}}{4}(x^{\mu}%
\Box+\Box x^{\mu}))  &  =\mathbf{E}(x^{\mu})[1+\left(  {k}_{\mu}[\frac
{(2\beta-\beta^{^{\prime}})\hbar^{2}}{4}(x^{\mu}\Box+\Box x^{\mu})]\right)
],\\
\mathbf{B}(x^{\mu}+\frac{(2\beta-\beta^{^{\prime}})\hbar^{2}}{4}(x^{\mu}%
\Box+\Box x^{\mu}))  &  =\mathbf{B}(x^{\mu})[1+\left(  {k}_{\mu}[\frac
{(2\beta-\beta^{^{\prime}})\hbar^{2}}{4}(x^{\mu}\Box+\Box x^{\mu})\right)  ],
\end{align}
Restrict to a one dimensional propagation we have
\begin{align}
\mathbf{E}(x^{\prime},t^{\prime})  &  =\mathbf{E}(x,t)[1+i\frac{(2\beta
-\beta^{^{\prime}})\hbar^{2}}{4}(kx-\omega t)\Box+i\frac{(2\beta
-\beta^{^{\prime}})\hbar^{2}}{4}\Box(kx-\omega t))],\\
\mathbf{B}(x^{\prime},t^{\prime})  &  =\mathbf{B}(x,t)[1+i\frac{(2\beta
-\beta^{^{\prime}})\hbar^{2}}{4}(kx-\omega t)\Box+i\frac{(2\beta
-\beta^{^{\prime}})\hbar^{2}}{4}\Box(kx-\omega t))],
\end{align}
As a consequence, the presence of a second order transformation in $\hbar$
implies a contribution resulting from a quantization, where the fields acts as
quantum observable.

\section{NCG transformation with $\beta^{\prime}=2\beta$ and modified Maxwell
equations by minimum length}

\subsection{Electromagnetic waves}

Considering electromagnetic waves in the presence of a minimum length, we have
the simplified form for these equations in the vacuum
\begin{align}
(1+\beta\hbar^{2}\Box)\nabla\cdot\mathbf{E}  &  =0\\
(1+\beta\hbar^{2}\Box)\nabla\cdot\mathbf{B}  &  =0,\\
(1+\beta\hbar^{2}\Box)\nabla\times\mathbf{E}  &  =-(1+\beta\hbar^{2}\Box
)\frac{\partial\mathbf{B}}{\partial t},\\
(1+\beta\hbar^{2}\Box)\nabla\times\mathbf{B}  &  =\mu_{0}\varepsilon
_{0}(1+\beta\hbar^{2}\Box)\frac{\partial\mathbf{E}}{\partial t}, \label{s214}%
\end{align}
Let us consider the defined fields
\begin{align}
\tilde{\mathbf{E}}  &  =(1+\beta\hbar^{2}\Box)\mathbf{E},\\
\tilde{\mathbf{B}}  &  =(1+\beta\hbar^{2}\Box)\mathbf{B},
\end{align}
We then have that for the linearly polarized solution for the electromagnetic
waves
\begin{align}
\mathbf{E}(x,t)  &  =\mathbf{E}_{0}\cos(kx-\omega t),\\
\mathbf{B}(x,t)  &  =\mathbf{B}_{0}\cos(kx-\omega t),\\
&
\end{align}
the modified fields will have the form
\begin{align}
\tilde{\mathbf{E}}(x,t)  &  =\mathbf{E}_{0}\cos(kx-\omega t)+\beta\hbar
^{2}(-k^{2}+\frac{\omega^{2}}{c^{2}})\mathbf{E}_{0}\cos(kx-\omega t),\\
\tilde{\mathbf{B}}(x,t)  &  =\mathbf{E}_{0}\cos(kx-\omega t)+\beta\hbar
^{2}(-k^{2}+\frac{\omega^{2}}{c^{2}})\mathbf{E}_{0}\cos(kx-\omega t),\\
&
\end{align}
As a consequence, the electromagnetic waves are left invariant in this case,
since the left-hand side of these equation vanishes when the electromagnetic
fields are wave solutions. This result, asserts that the non-commutativity
displayed does not affect electromagnetic waves.

\subsection{Source terms}

The presence of a minimum length transformation due to non-commutativity
(\ref{em3}) induces a changing in the Maxwell equations, particularly in the
Gauss and Amp\`{e}re-Maxwell laws.
The set of modified Maxwell equations in the presence of source terms is given
by
\begin{align}
(1+\beta\hbar^{2}\Box)\nabla\cdot\mathbf{E} &  =\frac{\rho}{\varepsilon_{0}%
},\label{s31}\\
(1+\beta\hbar^{2}\Box)\nabla\cdot\mathbf{B} &  =0,\\
(1+\beta\hbar^{2}\Box)\nabla\times\mathbf{E} &  =-(1+\beta\hbar^{2}\Box
)\frac{\partial\mathbf{B}}{\partial t},\\
(1+\beta\hbar^{2}\Box)\nabla\times\mathbf{B} &  =\mu_{0}\mathbf{J}+\mu
_{0}\varepsilon_{0}(1+\beta\hbar^{2}\Box)\frac{\partial\mathbf{E}}{\partial
t},\label{s34}%
\end{align}
As usual, let us split the current density in terms of free, polarization and
magnetization contributions $\mathbf{J}=\mathbf{J}_{f}+\mathbf{J}_{\mathbf{P}%
}+\mathbf{J}_{\mathbf{M}}$, and the charge density in terms of free and
polarization terms $\rho=\rho_{f}+\rho_{\mathbf{P}}$. The material media
contribution is associated to the presence of an electric polarization
$\mathbf{P}$ and a magnetization $\mathbf{M}$. These allows to write
\begin{align}
\mathbf{J} &  =\mathbf{J}_{f}+\frac{\partial\mathbf{P}}{\partial t}%
+\nabla\times\mathbf{M}\\
\rho &  =\rho_{f}-\nabla\cdot\mathbf{P}%
\end{align}
We can then rewrite the previous equations as
\begin{align}
\nabla\cdot\left[  (1+\beta\hbar^{2}\Box)\mathbf{E}+\frac{\mathbf{P}%
}{\varepsilon_{0}}\right]   &  =\frac{\rho_{f}}{\varepsilon_{0}}%
,\label{s31l}\\
\nabla\cdot\left[  (1+\beta\hbar^{2}\Box){\mathbf{B}}\right]   &  =0,\\
\nabla\times\left[  (1+\beta\hbar^{2}\Box)\mathbf{E}\right]   &
=-\frac{\partial\left[  (1+\beta\hbar^{2}\Box){\mathbf{B}}\right]  }{\partial
t},\\
\nabla\times\left[  (1+\beta\hbar^{2}\Box){\mathbf{B}}-\mu_{0}\mathbf{M}%
\right]   &  =\mu_{0}\mathbf{J}_{f}+\frac{\partial\left[  \mu_{0}%
\varepsilon_{0}(1+\beta\hbar^{2}\Box){\mathbf{E}}+\mu_{0}\mathbf{P}\right]
}{\partial t},\label{s34l}%
\end{align}
The corresponding fields in material media can be defined with $\beta$
dependence
\begin{align}
\mathbf{D}_{\beta} &  =\left(  \varepsilon_{0}(1+\beta\hbar^{2}\Box
)\mathbf{E}+\mathbf{P}\right)  \nonumber\\
&  =\mathbf{D}+\beta\hbar^{2}\Box\mathbf{E}%
\end{align}
where we also have a generalized response $\mathbf{H}$ to the material media
\begin{align}
\mathbf{H}_{\beta} &  =\frac{(1+\beta\hbar^{2}\Box)}{\mu_{0}}\mathbf{B}%
-\mathbf{M}\nonumber\\
&  =\mathbf{H}+\frac{\beta\hbar^{2}\Box}{\mu_{0}}\mathbf{B}.
\end{align}
We then have the set of modified equations rewritten as
\begin{align}
\nabla\cdot\mathbf{D}_{\beta} &  =\rho_{f},\label{s31lq}\\
\mu_{0}\nabla\cdot\left[  \mathbf{H}_{\beta}+\mathbf{M}\right]   &  =0,\\
\frac{1}{\varepsilon_{0}}\nabla\times\left[  \mathbf{D}_{\beta}-\mathbf{P}%
\right]   &  =-\mu_{0}\frac{\partial\left[  \mathbf{H}_{\beta}+\mathbf{M}%
\right]  }{\partial t},\\
\nabla\times\mathbf{H}_{\beta} &  =\mathbf{J}_{f}+\frac{\partial
\mathbf{D}_{\beta}}{\partial t},\label{s34lq}%
\end{align}
Using the constitutive relation for polarization $\mathbf{P}=\varepsilon
_{0}\chi_{e}\mathbf{E}$ and the generalized one for magnetization
\[
\mathbf{M}=\chi_{m,\beta}\mathbf{H}_{\beta},
\]
where $\chi_{m,\beta}$ is a $\beta$-dependent parameter. We can then write
generalized relations for the fields in material media
\begin{align}
\mathbf{D}_{\beta} &  =\varepsilon_{0}\left(  1+\beta\hbar^{2}\Box+\chi
_{e}\right)  \mathbf{E},\\
\mathbf{H}_{\beta} &  =\frac{1}{(1+\chi_{m,\beta})\mu_{0}}(1+\beta\hbar
^{2}\Box)\mathbf{B}.
\end{align}

\section{Metamaterials from modified Maxwell equations with minimum length}

In the Fourier transformed space in the previous result, we have
\begin{align}
\mathbf{D}_{\beta}(\mathbf{p},p_{0}=\omega)  &  = \varepsilon_{0}\left(
1+\beta\hbar^{2}p_{\mu}p^{\mu}+ \chi_{e}\right)  \mathbf{E}(\mathbf{p}%
,p_{0}=\omega),\\
\mathbf{H}_{\beta}(\mathbf{p},p_{0}=\omega)  &  = \frac{1}{(1+\chi_{m,\beta
})\mu_{0}}(1+\beta\hbar^{2}p_{\mu}p^{\mu})\mathbf{B}(\mathbf{p},p_{0}=\omega).
\end{align}
As a consequence, the material media will be identified by mean of material
pemissivity and permeabilities
\begin{align}
\varepsilon_{\beta}  &  = \varepsilon_{0}\left(  1+\beta\hbar^{2}p_{\mu}%
p^{\mu}+ \chi_{e}\right) \\
\mu_{\beta}  &  = \frac{(1+\chi_{m,\beta})\mu_{0}}{(1+\beta\hbar^{2}p_{\mu
}p^{\mu})}%
\end{align}
Taking case where $|\chi_{m,\beta}|>> 1$ and $\chi_{e} << 1$, we have the
refraction index given by a dependence in the generalized NCG parameter
\begin{align}
n_{\beta} = \sqrt{\chi_{m,\beta}}%
\end{align}
Considering $\chi_{m,\beta}$ given by a complex term
\begin{align}
\chi_{m,\beta}= \mathcal{T}_{\beta}e^{i\vartheta_{\beta}}%
\end{align}
we then have an negative index refraction metamaterial for $\vartheta_{\beta
}=2\pi$, where $e^{i\vartheta_{\beta}/2}=-1$, given by
\begin{align}
n_{\beta}  &  = -\sqrt{\mathcal{T}_{\beta}}.
\end{align}
This result shows that a metamaterial like behaviour can be achieved in the
presence of a non-commutative geometry with artificial control of the
non-commutative parameters in material media.

\section{Modified Maxwell equations from CPT-odd standard model extension and
minimum length}

The set of modified Maxwell equations in the CPT-odd standard model extension
and minimum length described above is given by
\begin{align}
(1+2\beta\hbar^{2}\Box)\nabla\cdot\mathbf{E} &  -c\chi\mathbf{V}\cdot
(1+2\beta\hbar^{2}\Box)\mathbf{B}=\frac{\rho}{\varepsilon_{0}}\\
\nabla\cdot\mathbf{B} &  =0,\\
\nabla\times\mathbf{E} &  =-\frac{\partial\mathbf{B}}{\partial t},\\
(1+2\beta\hbar^{2}\Box)\left(  \nabla\times\mathbf{B}-\frac{1}{c^{2}}%
\frac{\partial\mathbf{E}}{\partial t}\right)   &  -\chi V_{0}(1+\beta\hbar
^{2}\Box)\mathbf{B}+\chi\mathbf{V}\times(1+\beta\hbar^{2}\Box)\frac
{\mathbf{E}}{c}=\mu_{0}\mathbf{J}\nonumber\\
&
\end{align}
We can define field dependent generalized charge density and charge current,
given by
\[
\frac{\tilde{\rho}(\mathbf{B},\mathbf{V})}{\varepsilon_{0}}=\frac{\rho
}{\varepsilon_{0}}+c\chi\mathbf{V}\cdot(1+2\beta\hbar^{2}\Box)\mathbf{B}%
\]
and
\[
\mu_{0}\tilde{\mathbf{J}}(\mathbf{E},\mathbf{B},\mathbf{V},V_{0})=\chi
V_{0}(1+\beta\hbar^{2}\Box)\mathbf{B}-\chi\mathbf{V}\times(1+\beta\hbar
^{2}\Box)\frac{\mathbf{E}}{c}+\mu_{0}\mathbf{J}%
\]
The equations can the be rewritten as
\begin{align}
(1+2\beta\hbar^{2}\Box)\nabla\cdot\mathbf{E} &  =\frac{\tilde{\rho}%
(\mathbf{B},\mathbf{V})}{\varepsilon}\\
\nabla\cdot\mathbf{B} &  =0,\\
\nabla\times\mathbf{E} &  =-\frac{\partial\mathbf{B}}{\partial t},\\
(1+2\beta\hbar^{2}\Box)\left(  \nabla\times\mathbf{B}-\frac{1}{c^{2}}%
\frac{\partial\mathbf{E}}{\partial t}\right)   &  =\mu_{0}\tilde{\mathbf{J}%
}(\mathbf{E},\mathbf{B},\mathbf{V},V_{0})
\end{align}
We now split the current density in terms of free, polarization and
magnetization and LV contributions
\begin{align}
\mathbf{J} &  =\mathbf{J}_{f}+\mathbf{J}_{\mathbf{P}}+\mathbf{J}_{\mathbf{M}%
}+\mathbf{J}_{\mathbf{W}},\\
\rho &  =\rho_{f}+\rho_{\mathbf{P}}+\rho_{\mathbf{W}}%
\end{align}
where we have generalized current and charge density responses to the LV
background given by
\begin{align}
\mathbf{J}_{\mathbf{W}} &  =-\chi V_{0}\mathbf{M}-\frac{\chi}{\mu
_{0}\varepsilon_{0}}\mathbf{V}\times\frac{\mathbf{P}}{c}\\
\rho_{\mathbf{W}} &  =-\varepsilon_{0}\mu_{0}c\chi\mathbf{V}\cdot\mathbf{M}.
\end{align}
The generalized charge density can be written
\[
\frac{\tilde{\rho}(\mathbf{B},\mathbf{V})}{\varepsilon_{0}}=\frac{\rho
_{f}-\nabla\cdot\mathbf{P}}{\varepsilon_{0}}-\mu_{0}c\chi\mathbf{V}%
\cdot\mathbf{M}+c\chi\mathbf{V}\cdot(1+2\beta\hbar^{2}\Box)\mathbf{B}%
\]
We can then write
\[
\frac{\tilde{\rho}(\mathbf{B},\mathbf{V})}{\varepsilon_{0}}=\frac{\rho
_{f}-\nabla\cdot\mathbf{P}}{\varepsilon_{0}}+\mu_{0}c\chi\mathbf{V}%
\cdot\mathbf{H}_{\beta}.
\]
and the generalized current density is given by
\begin{align}
\mu_{0}\tilde{\mathbf{J}}(\mathbf{E},\mathbf{B},\mathbf{V},V_{0}) &  =\chi
V_{0}(1+\beta\hbar^{2}\Box)\mathbf{B}-\chi\mathbf{V}\times(1+\beta\hbar
^{2}\Box)\frac{\mathbf{E}}{c}\nonumber\\
&  +-\mu_{0}\chi V_{0}\mathbf{M}-\frac{\chi}{\varepsilon_{0}}\mathbf{V}%
\times\frac{\mathbf{P}}{c}+\mu_{0}\left(  \mathbf{J}_{f}+\frac{\partial
\mathbf{P}}{\partial t}+\nabla\times\mathbf{M}\right)
\end{align}
we can write it
\[
\mu_{0}\tilde{\mathbf{J}}(\mathbf{E},\mathbf{B},\mathbf{V},V_{0})=\mu_{0}\chi
V_{0}\mathbf{H}_{\beta}-\frac{\chi}{\varepsilon_{0}c}\mathbf{V}\times
\mathbf{D}_{\beta}+\mu_{0}\left(  \mathbf{J}_{f}+\frac{\partial\mathbf{P}%
}{\partial t}+\nabla\times\mathbf{M}\right)
\]
The Modified Maxwell equations in material media are then written in the form
\begin{align}
\nabla\cdot\mathbf{D}_{\beta} &  =\rho_{f}+\frac{\chi}{c}\mathbf{V}%
\cdot\mathbf{H}_{\beta},\label{s31lqq}\\
\nabla\cdot\left[  \mathbf{H}_{\beta}+\mathbf{M}\right]   &  =0,\\
\nabla\times\left[  \mathbf{D}_{\beta}-\mathbf{P}\right]   &  =-\frac{1}%
{c^{2}}\frac{\partial\left[  \mathbf{H}_{\beta}+\mathbf{M}\right]  }{\partial
t},\\
\nabla\times\mathbf{H}_{\beta} &  =\mathbf{J}_{f}+\frac{\partial
\mathbf{D}_{\beta}}{\partial t}+\chi V_{0}\mathbf{H}_{\beta}-c\chi
\mathbf{V}\times\mathbf{D}_{\beta},\label{s34lqq}%
\end{align}
We note that these set of equations include the presence of the vectorial
background due to Lorentz violation interacting with the material fields.
Additionally, the presence of non-commutativitiy is encapsulated in the fields
and can be used in the way to build a suitable metamaterial.
%

\section{Concluding remarks}

We have considered an standard model extension involving a CPT-odd sector in
the presence of a non-commutative geometry with minimum length. We considered
the cases of electromagnetic waves and material media. In particular, we
considered the case of metamaterial behaviour leading to the presence of a
negative index of refraction in dependent of the modified dielectric
contributions. We also derived a set of modified Maxwell equations in material
media, where the presence of the tensor background, as a result of Lorentz
violation, explicitly appears in the interaction terms with the material fields.

As the standard model extension can be used in this context to derive suitable
metamaterials with novel properties, we can consider this result as an
important step to determine the relationship between metamaterial behaviour,
in particular with negative index refraction, with the presence of
non-commutative geometry and Lorentz violation.

\textbf{Ackowledgements}

The authors acknowledge the supports by CNPq, CAPES, FAPES and
FAPEMA-UNIVERSAL-01401/16 (Brazil).

\end{document}